\documentclass[%
 reprint,
 superscriptaddress,
 amsmath,amssymb,
 aps,
prb,
floatfix,
]{revtex4-2}

\usepackage{graphicx}
\usepackage{dcolumn}
\usepackage{bm}
\usepackage{xcolor}
\usepackage{soul}

\usepackage{lipsum}

\usepackage{ifpdf,braket}
\ifpdf
\usepackage[pdftex,
        colorlinks=true,
        citecolor=blue,
        linkcolor=blue,
        urlcolor=blue,
        pdftoolbar=true,
        pdfmenubar=false,
        pdfpagemode=UseOutlines,
        pdfstartview=FitH,
        bookmarksopen=false
        ]{hyperref}
\fi

\begin{document}

\preprint{APS/123-QED}

\title{Closed vortex state in 3D mesoscopic superconducting films under an applied transport current}

\author{Leonardo Rodrigues Cadorim}
\affiliation{%
Departamento de F\'isica, Faculdade de Ci\^encias, 
Universidade Estadual Paulista (UNESP), Caixa Postal 473, 
17033-360, Bauru-SP, Brazil
}%
\affiliation{%
Departement Fysica, Universiteit Antwerpen, Groenenborgelaan 171, B-2020 Antwerp, Belgium
}%


\author{Lucas Veneziani de Toledo}
\affiliation{%
Departamento de F\'isica, Faculdade de Ci\^encias, 
Universidade Estadual Paulista (UNESP), Caixa Postal 473, 
17033-360, Bauru-SP, Brazil
}%

\author{Wilson Aires Ortiz}
\affiliation{%
Departamento de F\'isica,  
Universidade Federal de S\~ao Carlos (UFSCar), Caixa Postal 676, 
13565-905, S\~ao Carlos-SP, Brazil
}%

\author{Jorge Berger}
\affiliation{%
Department of Physics and Optical Engineering, Braude 
College, Karmiel, 2161002, Israel
}%

\author{Edson Sardella}
\email{Corresponding author: edson.sardella@unesp.br}
\affiliation{%
Departamento de F\'isica, Faculdade de Ci\^encias, 
Universidade Estadual Paulista (UNESP), Caixa Postal 473, 
17033-360, Bauru-SP, Brazil
}%


\date{\today}

\begin{abstract}

By using the full 3D generalized time dependent Ginzbug-Landau equation 
we study a long superconducting film of finite width and thickness 
under an applied transport current. We show that, for 
sufficiently large thickness, the vortices and the antivortices 
become curved before they annihilate each 
other. As they approach the center of the sample, 
their ends combine, producing a single closed vortex. 
We also determine the critical values of the thickness 
for which the closed vortex sets in for different values of 
the Ginzburg-Ladau parameter. Finally, we propose a 
model of how to detect a closed vortex experimentally.
\end{abstract}

\maketitle


\section{\label{sec:sec1}Introduction}



One of the most outstanding physical phenomena in condensed 
matter theory is the flux quantization in the Shubnikov phase 
of superconducting materials. In bulk type-II superconductors,
above a certain critical value of applied field, magnetic flux 
penetrates the sample in the form of flux quanta 
(in units of $\Phi_0=hc/2e$), 
the so called magnetic vortices. Superconducting vortices are 
very important due to their possible use in 
many applications, ranging from single photon detectors 
to quantum information. Moreover, the knowledge of how 
vortices behave under various circumstances 
is fundamental to the optimization of many electronic devices, since
the vortex  motion causes energy 
dissipation in a superconductor.

In the core of a superconducting vortex, the order 
parameter that describes the superconducting state 
vanishes and its phase changes by a multiple 
of $2\pi$ when circulated around a closed loop that encloses the core.
As the vortex line tends to align with the direction of the magnetic 
field, in the presence of an external field applied perpendicularly to a 
superconducting film, the core of 
the vortex is rigorously a straight line with the 
currents flowing around it. This is the so called Abrikosov vortex, 
first described in his outstanding work \cite{abrikosov56}.

However, the circularly shaped magnetic self-field induced by a
current produces an Abrikosov vortex with a 
ringlike shape, which is usually called a 
\textit{closed vortex}. This new type of solution 
was first found by 
Kozlov and Samokhvalov \cite{kozlov1991} 
through the solution of the London equation
and was extensively studied further in 
Refs.~\cite{KOZLOV1993103,kozlov1993,GENENKO1993343,genenko1994vortex,PhysRevB.49.6950,PhysRevB.51.3686,PhysRevB.57.1164,SAMOKHVALOV1996337,SAMOKHVALOV19972163,SAMOKHVALOV199874}. 
In these works, the existence, dynamics and even 
stability of such closed vortices in the presence of 
inhomogeneities were studied for unbounded superconductors 
or superconducting samples in the shape of a cylinder. 
In the latter case, due to matching symmetry between 
the field lines produced by the current and the cylinder 
geometry, the vortex penetrates the superconductor already 
in the form of a closed ring. Recently, these closed vortices 
were also shown to exist in a more complex geometry, such as 
a superconducting torus. \cite{niedzielski2022}

In different scenarios, the formation of closed vortices was also 
recently studied, for example, in Josephson junctions \cite{berdiyorov2018} 
submitted to an external current, where it was shown that Josephson vortices 
can also be found in the form of closed vortex loops and a procedure 
to their experimental observation was introduced. Due to short life 
time of the Abrikosov closed vortices, their experimental detection 
is a difficult task and has not been accomplished so far. Recently, by 
using the microscopic theory of superconductivity Fyhn and 
Linder \cite{fyhn2019} proposed an experimental setup to 
the observation of such objects based on STM measurements. One of 
the most important results of the present work is to provide a 
new procedure to the detection of closed vortex loops, which 
is based on the measurement of the magnetic field profile produced by them.
It is worth mentioning that, in practice, half-closed vortices have been demonstrated in RF cavities. \cite{gurevich2017}

In the present work, we study the physics of the closed 
vortices in a different setup. Namely, we investigate 
how a closed vortex emerges and gets annihilated 
in a superconducting slab under the presence of a transport 
current.
Unlike previous works on the subject, the geometry of our superconductor does not 
favor the formation of a closed vortex. Here, this object is created 
solely by the inhomogeneous action of the applied current 
in different parts of the flux line.
The study of the vortex dynamics is of fundamental importance to the 
understanding of the resistive state of current driven 
superconductors. Here, we show that, for appropriate 
thickness of the slab and Ginzburg-Landau 
parameter $\kappa$, a closed vortex forms in the process 
of annihilation of a vortex-antivortex (v-av) pairs induced in 
the sample by the current self-field. As we show, the  
lines corresponding to the vortex and the antivortex combine 
and form a closed loop when the pair is annihilating. After 
this combination, the loop takes the form of a quasi-ellipse, with 
the aspect ratio gradually decreasing until the collapse of the loop. 
By increasing the thickness of the film, the loop 
tends to a circle.

The outline of this work is as follows. In Sec.~\ref{sec:sec2} 
we present our model and the formalism we used in order 
to solve the generalized Ginzburg-Landau equations 
\cite{kramer1978theory,watts1981nonequilibrium}. In Sec.~\ref{sec:sec3} 
we present and discuss the results obtained in our simulations.
Finally, we present our concluding remarks in Sec.~\ref{sec:sec4}.
    
\section{\label{sec:sec2} Theoretical model}

In this work, we rely on the generalized time dependent 
Ginzburg-Landau (GTDGL) equation which is more suitable 
to describe the resitive state of dirty superconductors in the
non-equilibrium state \cite{kramer1978theory,watts1981nonequilibrium}.  
In dimensionless units this equation is given by
\begin{eqnarray}
    \frac{u}{\sqrt{1+\gamma^2|\psi|^2}}\left [ \frac{\partial }{\partial t}
    +\frac{1}{2}\gamma^2\frac{\partial |\psi|^2}{\partial t} \right ] \psi = & & \nonumber \\ 
    = \left(\mbox{\boldmath $\nabla$}-i\textbf{A}\right)^{2}\psi+\psi(1-|\psi|^2),& & 
    \label{eq:eq1}
\end{eqnarray}
coupled with Ampère's law
\begin{equation}
    \Sigma \frac{\partial \textbf{A}}{\partial t} 
    = \textbf{J}_s-\kappa^2\mbox{\boldmath $\nabla$}\times\textbf{h},
    \label{eq:eq2}
\end{equation}
where 
\begin{equation}
\textbf{J}_s = {\rm Im}\left [\bar{\psi}{(\mbox{\boldmath $\nabla$}-i\textbf{A})}\psi\right]
\label{eq:eq3}
\end{equation}
is the superconducting current density. 

Here, the temperature is in units of the critical temperature $T_c$; 
the order parameter $\psi$ is in units of $\psi_\infty(T)=\sqrt{\alpha(T)/\beta}$, 
where $\alpha$ and $\beta$ are two phenomenological constants; 
the distances are measured in units of the coherence length $\xi(T)$; 
the vector potential $\textbf{A}$ is in units of $\xi H_{c2}(T)$, where $H_{c2}$ 
is the upper critical field; the local magnetic field 
$\textbf{h}=\mbox{\boldmath $\nabla$}\times\textbf{A}$ is units of $H_{c2}(T)$; 
time is in units of the Ginzburg-Landau characteristic 
time $\tau_{GL}=\pi\hbar/8k_BTu$; the material 
dependent parameter $\gamma=2\tau_E\Delta_0/\hbar$, 
where $\tau_E$ is the inelastic electron-collision time, and 
$\Delta_0$ is the gap in the Meissner state; 
the constant $\Sigma=4\pi\sigma D/c^2\xi^2(T)$, where 
$D$ is the diffusion coefficient and $\sigma$ is 
the normal state electrical conductivity; $\kappa=\lambda(T)/\xi(T)$ is 
the Ginzburg-Landau parameter, where $\lambda(T)$ is 
the London penetration depth; 
and finally, the constant $u$ 
is equal to 5.79, which is derived from first 
principles \cite{kramer1978theory}.

The original GTDGL equations take into account the 
scalar electrical potential $\varphi$. Since they are invariant 
under the following gauge transformations
\begin{eqnarray}
    \psi^\prime       & = & e^{-i\chi}\psi, \nonumber \\
    \textbf{A}^\prime & = & \textbf{A}-\mbox{\boldmath $\nabla$}\chi, \nonumber \\
    \varphi^\prime    & = & \varphi+\frac{\partial\chi}{\partial t}, \label{eq:eq4}
\end{eqnarray}
where $\chi$ is an arbitrary scalar function 
therefore, we conveniently use the Weyl gauge \cite{du1996high} 
in which the scalar potential is constant and equal to zero.

\begin{figure*}[htpb]
    \centering
    \includegraphics[width=0.98\textwidth]{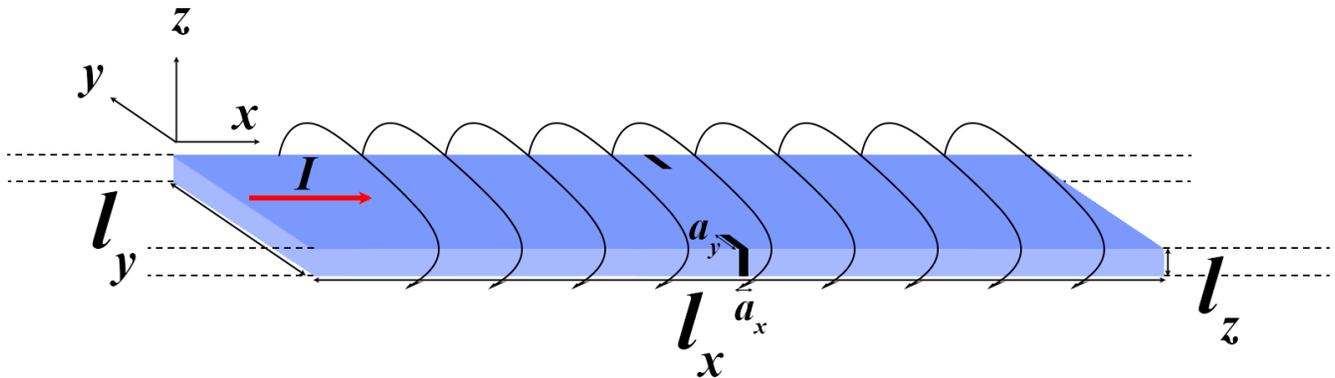}
    \caption{Schematic view of the system under consideration: an 
    infinitely long superconducting sample of width $l_y$ and thickness 
    $l_z$; only one unit cell of length $l_x$ is shown. The 
    transport current is applied in the $x$ direction. The encircling 
    lines illustrate the line fields of the self-field produced by 
    the current. Two defects are introduced at the border 
    of the sample (black spots), in order to facilitate nucleation 
    of v-av pairs.}
    \label{fig:fig1}
\end{figure*}

In the present work, we consider an infinite superconducting 
film carrying a transport current as sketched in Fig.~\ref{fig:fig1}. 
The applied transport current is introduced as follows. 
The superconductor is in the presence of an applied 
electric field which is sustained by a DC transport current density 
which flows along the $x$ axis, $\textbf{J}_a=J_a\hat{\textbf{x}}$ .
In the normal 
state, the vector potential has only the $x$ component. Thus, we  
separate the vector potential and the local magnetic field into two contributions, 
one coming from the normal state, and another one due to the 
diamagnetic nature of the superconductor that tends to cancel out 
the field induced by the applied current inside the 
sample. 
In other words, in Eqs.\ref{eq:eq1}-\ref{eq:eq3} we substitute:
\begin{eqnarray}
    \textbf{A} & = & \textbf{A}_0+\textbf{A}_1, \label{eq:eq7}\\
    \textbf{h} & = & \textbf{h}_0+\textbf{h}_1, \label{eq:eq8}
\end{eqnarray}
where \textbf{A}$_0$  and \textbf{h}$_0$ 
satisfy the following equations:
\begin{eqnarray}
    \kappa^2\mbox{\boldmath $\nabla$}\times\textbf{h}_0 = J_a\hat{\textbf{x}},\;\;\;
    \kappa^2\nabla^2 A_{0x} = -J_a\,.
    \label{eq:eq9}
\end{eqnarray}

The analytical solutions of equations (\ref{eq:eq9}) are given in the 
Supplementary Material of Ref.~\cite{cadorim2020ultra}.

Here, we solve the full 3D GTDGL equations 
numerically for an infinite superconducting 
film of finite width and thickness (see Fig.~\ref{fig:fig1}). 
The infinitely long film is divided into unit cells of dimensions $(l_x,l_y,l_z)$. We  
take into account the demagnetization effects. Therefore, for numerical 
purposes, we must consider the unit cell inside a 
simulation box (not shown in Fig.~\ref{fig:fig1}) 
of dimensions $(l_x,L_y,L_z)$, where $(L_y,L_z)$ are 
sufficiently larger than $(l_y,l_z)$ so that the demagnetizing 
field $\textbf{h}_1$ vanishes far away from the superconducting 
surfaces (for more details see Ref.~\cite{barba2015superconducting}). 
In addition, the normal components of the 
superconducting current density of 
must be zero on the superconductor-vacuum interfaces.  
Then, the following boundary conditions must be fulfilled:
\begin{eqnarray}
    \hat{\textbf{n}}\cdot(\mbox{\boldmath $\nabla$}-i\textbf{A}_0-i\textbf{A}_1)\psi 
    & = & 0,\;\;\;\textrm{in}\; \partial\Omega_{sc}, \label{eq:eq13} \\
    \mbox{\boldmath $\nabla$}\times\textbf{A}_1 & = & 0,\;\;\;\textrm{in}\; \partial\Omega,
    \label{eq:eq14}
\end{eqnarray}
where $\partial\Omega_{sc}$ and $\partial\Omega$ stand for 
superconducting and simulation box surfaces, respectively.

The two black spots in Fig.~\ref{fig:fig1} represent 
two defects on the border of the sample. They are 
introduced as an artifact to create 
an inhomogeneity in the current, that induces the vortices and antivortices 
in the opposite sides of the sample.

\section{\label{sec:sec3}Results and discussion}

    \subsection{Parameters and Methodology}

    The above equations are discretized by using 
    the standard link-variable method as described in 
    Ref.~\cite{gropp1996numerical}. 
    This algorithm is then implemented in Fortran 90 programming 
    language and run in a 
    GPU (Graphics Processing Unit) 
    accelerated forward-time-central-space scheme.
    
    In the simulations, we have 
    fixed some parameters and varied others as follows. 
    The length and width of the unit cell are fixed as $l_x=12\xi$ and 
    $l_y=8\xi$. The thickness of the sample varied from $l_z=1\xi$ to
    $l_z=3.6\xi$ in increments of $0.2\xi$. 
    We use $\kappa=1/\sqrt{2},1,\sqrt{3}$ 
    for each set of values of $l_z$. The grid space used is 
    $\Delta x=\Delta y =0.2\xi$ 
    and $\Delta z=0.1\xi$. The size of the simulation box 
    was chosen sufficiently large in order to 
    satisfy boundary conditions (\ref{eq:eq14}); 
    we use $l_y=16\xi$ and $l_z=12\xi$. The dimensions of the defects are 
    $a_x=a_y=0.2\xi$. 
    The range $10\le \gamma \le 20$ is suitable 
    for most metals like 
    Nb \cite{kramer1978theory,watts1981nonequilibrium,berdiyorov2009}; 
    we used $\gamma=10$. 

    Let us explain how we
    calculate the IV (current-voltage) and IR 
    (current-resistance) characteristics, which  
    are the measurable quantities in the resistive state. 
    In the Weyl gauge, the electrical field 
    is given by $\textbf{E}=-\partial \textbf{A}/\partial t$, 
    so, assuming that the voltage is measured between 
    electrodes at $z=0$ that cover the width of the film,
    the voltage across a unit cell is:
    \begin{eqnarray}
        U(t) & = & -\frac{1}{n_y-1}\sum_{j=2}^{n_y}\int_{-l_x/2}^{l_x/2}\,E_x(x,y_j,0) dx \nonumber \\ 
        & = & \frac{1}{n_y-1}\sum_{j=2}^{n_y}\int_{-l_x/2}^{l_x/2}\frac{\partial A_x(x,y_j,0)}{\partial t} dx,
        \label{eq:eq15}
    \end{eqnarray}
    where $n_y=l_y/\Delta y$, and $y_j=(j-n_y/2-1)\Delta y$ for 
    are the $y$ coordinates 
    of the mesh points.  
    The voltage is then calculated as a time average of $U(t)$. We have:
    \begin{equation}
        V = \frac{1}{\cal T}\,\int_0^{\cal T}\,U(t)\,dt, \label{eq:eq16}
    \end{equation}
    where ${\cal T}$ is the time corresponding to an appropriate number of 
    oscillations of $U(t)$.
    
    The applied current density was adiabatically increased in 
    steps of $\Delta J_a = 0.01J_{GL}$ from the Meissner state 
    until the superconductivity was fully destroyed. In the resistive 
    state, we moved from value of $J_a$ to $J_a+\Delta J_a$ only
    after the voltage $U(t)$ became periodic, which is 
    the same periodicity with which the v-av pairs are 
    formed and annihilated. When 
    multiple nucleations of v-av (vortex-antivortex) pairs are present, the voltage 
    looses its periodicity, and therefore we change the value of $J_a$ 
    only after 220 oscillations of $U(t)$ in order to obtain 
    a more accurate value of the time average voltage. 
    
    The results of all simulations are compiled in the 
    following Subsections.

    \subsection{Field Profile, Closed Vortex, and Current Distribution}

    \begin{figure}[htpb]
        \centering
        \includegraphics[angle=-90,width=0.45\textwidth]{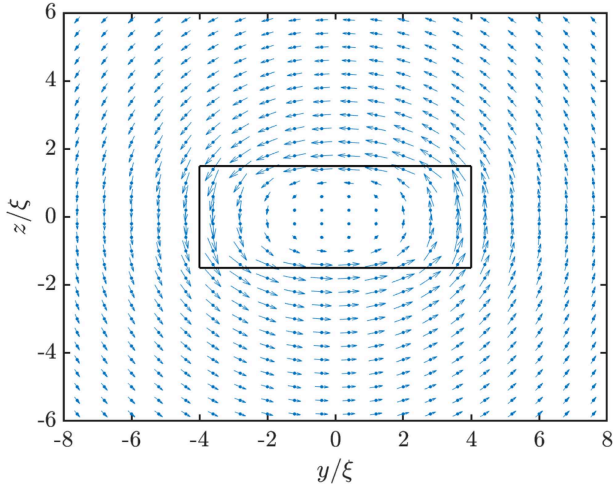}
        \caption{The magnetic field profile in the vertical plane $x=0$ 
        (parallel to the $yz$ plane): 
        for better visualization purposes, the arrows 
        are not in real size; the rectangle 
        inside is a cross section of the superconductor; 
        this picture is for $\kappa=1$, $l_y=8\xi$, 
        $l_z=3\xi$; the value of the current 
        density is $J_a=0.26J_{GL}$ just before 
        the critical current density $J_{c1}=0.27J_{GL}$. 
        The vortex (antivortex) nucleates on the right-hand side 
        (left hand-side) of the figure.}
        \label{fig:fig2}
    \end{figure}
    
    In Fig.~\ref{fig:fig2} 
    we illustrate the vector field profile 
    at $x=0$ plane (in the middle of 
    the unit cell, where the defects are located).
    The local field has symmetry 
    as if the current density was uniform. 
    Due to the geometric symmetry and 
    the demagnetization effect, the local magnetic 
    field is larger near the surface of the superconductor, 
    and decreases deep inside. As can also be 
    observed, the field is larger on the lateral sides of the sample. 
    In this figure, we show the vector field 
    for a value of the current 
    density just before the first critical current $J_{c1}=0.27J_{GL}$.
    Therefore, once the resistive state sets in, it is on the 
    lateral sides that the vortex and the antivortex  
    sprout, move to the center, and finally annihilate 
    each other at the center of the sample. Then, a periodic 
    collapsing of v-av pairs is established. 
    
    \begin{figure}[htpb]
        \centering
        \includegraphics[width=0.45\textwidth]{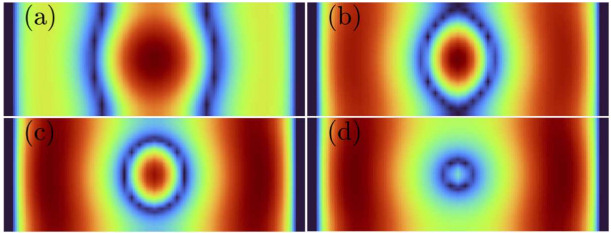}
        \caption{Color maps of the superconducting Cooper-pair 
        density, $|\psi|$,  for 
        $\kappa=1$, $l_z=3\xi$, and $J_a=0.27J_{GL}$ throughout the 
        plane $x=0$: (a) 
        a v-av pair of curved vortices; on the left hand-side 
        (right hand side) is the antivortex (vortex); 
        (b) a combination of a vortex 
        and an antivortex producing a closed vortex; (c) a 
        closed vortex diminishing its radius; 
        (d) a closed vortex 
        shrinking down at the center. The dark strips on 
        both sides are due 
        to the defects. These pictures correspond 
        to the same region highlighted 
        in Fig.~\ref{fig:fig2}.}
        \label{fig:fig3}
    \end{figure}
    
    Next, we discuss the morphology of a closed 
    vortex in the resistive state. The first works 
    about closed vortices were conducted on long current-carrying
    superconducting cylinders, so that  the vortex follows 
    the geometry of the sample since from the surface until 
    it collapses at the center \cite{kozlov1993,PhysRevB.49.6950}. 
    In the present scenario, 
    we deal with a film of rectangular cross section. 
    Thus, before the closed vortex is formed, 
    two curved vortices (a vortex and an antivortex) 
    nucleate in opposite sides of the sample 
    (see panel (a) of Fig.~\ref{fig:fig3}) 
    and move toward the center. Then, 
    as they encounter each 
    other, their ends join together forming a 
    closed vortex (panel (b)). Once this ringlike 
    vortex is formed, its radius starts decreasing 
    (panel (c)) until it collapses at the center. 
    After the transition from the Meissner state to 
    the resistive one, the process is repeated 
    periodically until superconductivity is suppressed 
    throughout the sample.
    
    \begin{figure}[htpb]
        \centering
        \includegraphics[width=0.49\textwidth]{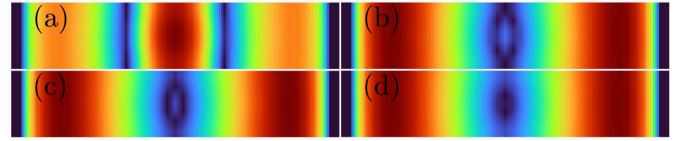}
        \caption{The same as in Fig.\ref{fig:fig3} for $l_z=1.6\xi$ and $J_a=0.30J_{GL}$.}
        \label{fig:fig4}
    \end{figure}
    
    It is conceivable that the closed vortex can exist for any 
    thickness of the film, but for small values of $l_z$ 
    is very elongated when the ends of the v-av pair 
    meet. In our simulations, we do not 
    have sufficient resolution to detect a closed vortex 
    for any $l_z$. Indeed, in Fig.~\ref{fig:fig4} we show four panels 
    of the color maps of the superconducting Cooper-pair density 
    for $l_z=1.6\xi$. As we can see, the shape 
    of the closed vortex is much more elongated than for the previous 
    case $l_z=3\xi$ of Fig.~\ref{fig:fig3}. For $\kappa=1$, 
    and thickness below $l_z=1.6\xi$, we do not observe 
    any closed vortex; the v-av pair remains straight lines, since 
    the nucleation of the vortex and the antivortex on the surfaces, 
    until the pair is annihilated at the center of the sample. The 
    formation of a closed vortex depends on the sample thickness 
    because the current and the magnetic field concentrate near the surface 
    of the superconductor, on a scale of the order of the London 
    penetration depth. Therefore, for thicker samples, the ends of 
    a vortex line are subjected to stronger Lorentz force and to a 
    stronger horizontal magnetic field that facilitate the formation 
    of a closed vortex.
    
    \begin{figure}[htpb]
        \centering
        \includegraphics[width=0.49\textwidth]{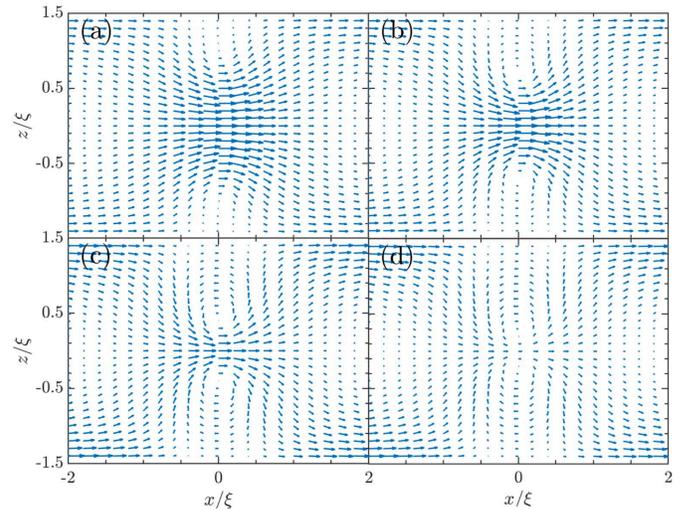}
        \caption{The panels show four cuts of the current 
        distribution of the closed vortex throughout 
        the vertical plane $y=0$. The radius of the closed vortex 
        diminishes from (a) to (d).}
        \label{fig:fig5}
    \end{figure}
    
    Let us now discuss the current distribution of a 
    closed vortex. When both curved vortex and antivortex 
    touch their ends on the upper and lower surfaces, 
    $z=+l_z/2$ and $z=-l_z/2$, respectively,  they 
    combine in order to make a single closed vortex. 
    This new vortex looks like a toroid with 
    the superconducting currents flowing around its core. 
    Fig.~\ref{fig:fig5} exhibits four 
    cuts of the toroid in the $xz$ plane for 
    $l_z=3\xi$. As it can be seen, the currents in 
    the internal parts of the toroid flow 
    in the same direction for both 
    the upper and lower segment of the closed 
    vortex. Therefore, all v-av pairs opposedly 
    positioned in the toroid attract one 
    another, causing the closed vortex to collapse 
    at the center of the sample.
    
    \subsection{Straight to Curved Vortex Crossover, and (IV,IR) Characteristics}
    
    \begin{figure}[htpb]
        \centering
        \includegraphics[width=0.49\textwidth]{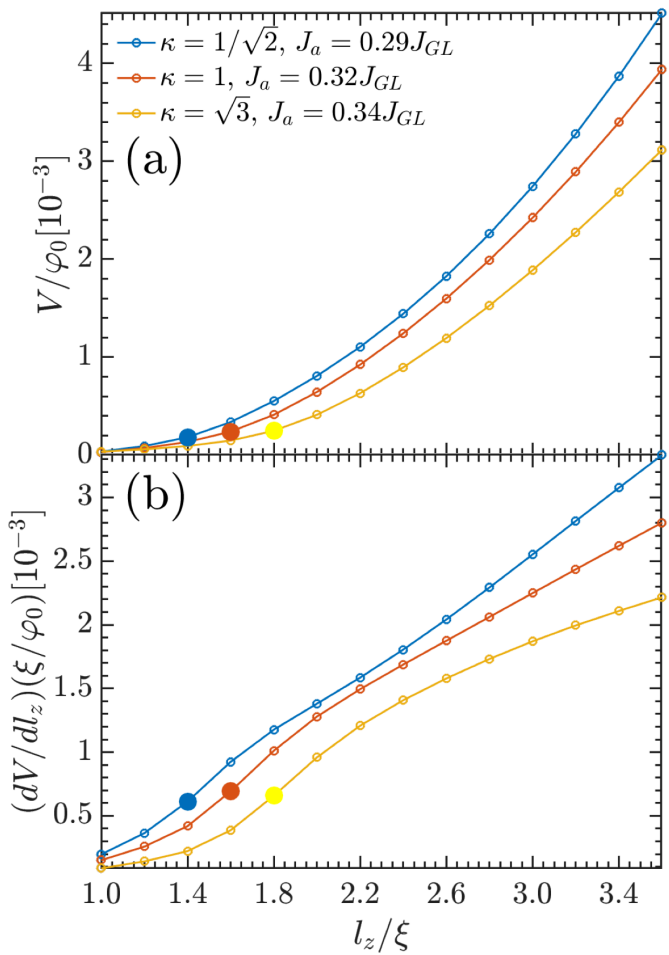}
        \caption{(a) Voltage across the $z$ direction as a function of 
        the thickness of the sample for three values of $\kappa$: the value 
        of $J_a$ for each case corresponds to the first critical 
        current density when the resistive state sets in. The highlighted
        dots are the critical $l_{z,c}$ values for which the v-av pairs 
        combine to make a closed vortex. (b) The derivative of 
        the voltage: the dots separate the two regimes of straight to 
        curved vortices; the inflection points coincide with $l_{z,c}$. 
        }
        \label{fig:fig6}
    \end{figure}
    
    As we mentioned previously, as the thickness of the sample is 
    increased, there is a crossover 
    between straight to curved v-av pair. In what follows, we show a 
    consistence between the criterion based on the 
    aspect ratio of the vortex (antivortex) and an important 
    physical quantity, namely, the voltage across the $z$ 
    direction on the lateral side of the film; by \textit{aspect ratio}, 
    we mean the distance between the center of the vortex and 
    the antivortex along the $y$ direction when their tips 
    first touch each other. For this 
    purpose, we calculate the time average of the following 
    voltage:
    \begin{eqnarray}
        U(t) & = & -\frac{2}{n_x}\sum_{i=n_x/2}^{n_x}\int_{0}^{l_z/2}\,E_z(x_i,l_y/2,z) dz \nonumber \\ 
        & = & \frac{2}{n_x}\sum_{i=n_x/2}^{n_x}\int_{0}^{l_z/2}\frac{\partial A_z(x_i,l_y/2,z)}{\partial t} dz,
        \label{eq:eq17}
    \end{eqnarray}
    where $n_x=l_x/\Delta x$, and $x_i=(i-n_x/2-1)\Delta x$ for all  
    $\{i = 1,2,\ldots ,n_x+1\}$ are the $x$ coordinates 
    of the mesh points. Here, we have not considered the branch 
    $-l_z/2 \le z \le 0$. By symmetry, had we included this 
    contribution, the voltage would vanish.
    
    When a closed vortex appears, we will have a larger 
    contribution for the current flowing 
    in the vertical direction, and consequently an 
    increase in the voltage. For a fixed value of 
    $\kappa$ and current density $J_a$, we determine 
    the voltage for several values of $l_z$. We have done 
    this for three distinct values 
    of the Ginzburg-Landau parameter. 
    The respective value of $J_a$  
    is chosen so as to correspond to the critical 
    current density $J_{c1}$ for the lowest thickness, 
    $l_z=1\xi$. The results are summarized in 
    Fig.~\ref{fig:fig6}. In panel (a) we 
    present the voltage as a function of the thickness 
    of the film. We find that, at a certain 
    point, which we denote by $l_{z,c}$, there is a change of the 
    behavior of the $V(l_z)$ curves. These points 
    are highlighted in panel (a). They signal 
    a crossover from linear to curved v-av pairs.
    
    In order to make sure that this special point 
    is correlated to the straight-to-curved vortex 
    crossover, we calculate the derivative 
    $dV(l_z)/dl_z$ for the three values of $\kappa$ 
    (see panel (b)).
    As can be seen, the derivatives have 
    an inflection point which are highlighted 
    in panel (b). These points correspond 
    to $l_{z,c}$.
    We find the following critical values, 
    $l_{z,c}=1.4\xi,1.6\xi,1.8\xi$ for 
    $\kappa=1/\sqrt{2},1,\sqrt{3}$, respectively. We must 
    emphasize that, first we determine the value of $l_{z,c}$ 
    by inspecting the aspect ratio of the curvature of the vortex. 
    Second, we check if the result is in agreement with 
    inflection point of $dV(l_z)/dl_z$.
    For all the three cases mentioned above 
    they coincide.
    
    \begin{figure}[htpb]
        \vspace{0.5cm}
        \centering
        \includegraphics[width=0.45\textwidth]{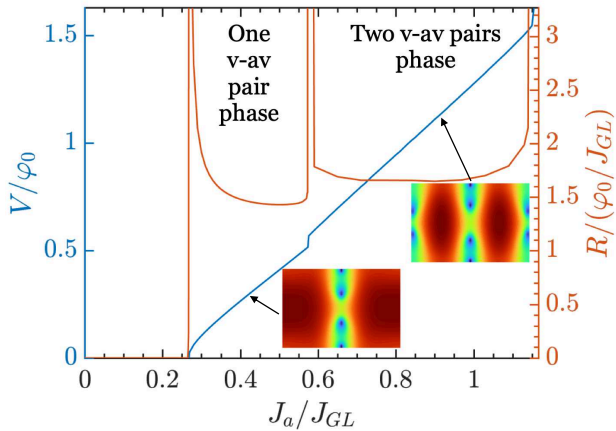}
        \caption{IV (blue line) and IR (red line) 
        characteristic curves, respectively, for $\kappa=1$ and $l_z=3\xi$. 
        The Meissner state (full superconductivity) survives up to 
        $J_a=J_{c1}=0.27J_{GL}$. Above this current density, 
        the resistive state sets in. The resistive state splits into two phases. 
        In one of them the vortex and the 
        antivortex nucleate only at the defects on the border of 
        the superconductor. 
        In the second phase, another 
        set of v-av pairs nucleates at 
        the frontiers between unit cells. 
        The second jump 
        in the IV characteristic is the signature of 
        this crossover. The insets illustrate this 
        scenario through the modulus of the order 
        parameter in the $xy$ plane ($z=0$ plane).
        }
        \label{fig:fig7}
    \end{figure}
    
    Now we discuss the transport properties 
    of the superconductor. The IV and IR characteristics are 
    presented in Fig.~\ref{fig:fig7} for 
    $\kappa=1$ and $l_z=3\xi$. As 
    we increase the applied current density, the system 
    becomes unstable to the penetration of v-av 
    pairs. When $J_a$ achieves the value $J_{c1}=0.27J_{GL}$ 
    the superconductor goes to the resistive state, 
    where a periodic 
    formation v-av pairs occurs. 
    Notice that $J_{c1}$ is 
    smaller than the depairing current density $J_{GL}$. 
    This is a consequence of the defects 
    deliberately introduced at the border 
    of the superconducting film. If we further increase 
    $J_a$, a second jump appears in the IV curve 
    at $J_a=0.575J_{GL}$. 
    This is an indication that another two 
    adjacent v-av pairs around the central one 
    are nucleating (see insets). This is in correspondence 
    with the experimental observations of 
    multiple penetrations of kinematic vortices 
    in Sn film by Sivakov \textit{et al.} \cite{Sivakov2003}. 
    Finally, when the current density reaches the value 
    $J_a=J_{c2}=1.155J_{GL}$ the superconductor goes straight 
    to the normal state. We believe that for larger unit cells, 
    additional jumps in the IV curve 
    would occur. 

    \subsection{Single Defect (Half-Closed Vortex)}    

    \begin{figure}[htpb]
        \vspace{0.5cm}
        \centering
        \includegraphics[width=0.49\textwidth]{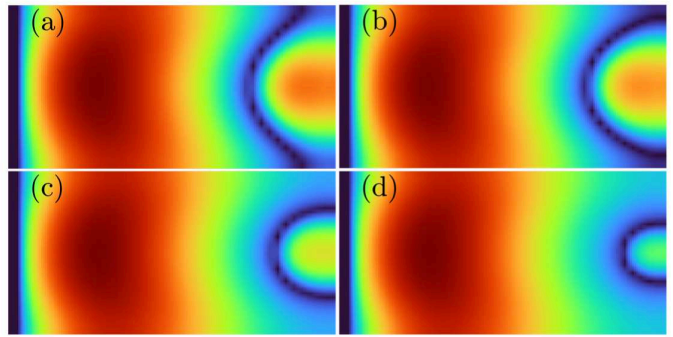}
        \caption{Color maps of the superconducting Cooper-pair 
        density, $|\psi|$,  for 
        $\kappa=1$, $l_z=4\xi$, and $J_a=0.26J_{GL}$ throughout the 
        plane $x=0$: (a) 
        an av nucleates in the left edge of the sample and moves towards 
        the opposite side;  
        (b) the ends of the av touch the $y=l_y/2$ plane and form a half-closed vortex; (c) a 
        half-closed vortex diminishing its radius; 
        (d) the half-closed vortex 
        shrinking down. 
        }
        \label{fig:fig8}
    \end{figure}
    
    We also have considered a single defect in the middle of 
    a unit cell (in the middle of front edge in Fig.~\ref{fig:fig1}). 
    In this configuration, only an antivortex nucleates 
    on the $y=-l_y/2$ surface. Once the antivortex nucleates at 
    $y = -l_y/2$ it moves directly towards the opposite side. 
    As can be seen from Fig.~\ref{fig:fig8}, as the antivortex approaches the 
    other side of the sample, it becomes significantly curved (see panel (a)). 
    When it reaches the surface $y=l_y/2$, surprisingly, 
    it does not escape the sample. Instead, its ends touch 
    the surface giving rise to a \textit{half-closed vortex} (see panels (b) and (c)). 
    Then, it diminishes its ratio until it collapses (see panel (d)).

    Due to its intrinsic nature, it is 
    very difficult to observe experimentally the 
    closed vortex. 
    In addition, both the closed 
    and half-closed vortex are very unstable. Therefore, we require 
    an indirect method that signals either a v-av 
    or a single half-closed vortex curves. 
    Having this in 
    mind, we propose a setup to detect 
    the curvature of the v-av pair when it  
    gives rise to a closed vortex 
    with a non vanishing aspect ratio.
    Instead of doing this for a closed vortex, which is formed inside the sample, 
    we think it should be much easier for a half-closed vortex, since 
    its collapse occurs on the surface. 
    For this purpose, we calculate 
    the time average of the of the magnetic flux 
    on the lateral side of the film. 
    In order to calculate the magnetic flux, we focus in a small region where 
    the antivortex tips touch the plane $y=l_y/2$, although we could 
    extend it throughout the whole lateral side of the unit cell. 
    We evaluate the following equation:
    \begin{equation}
        \Phi(t) = \int_{0}^{l_z/2} \int_{-\xi}^{\xi}\,h_y(x,\pm l_y/2,z)\,dx dz. \label{eq:eq18}
    \end{equation}
    Here, the minus sign stands for the left edge where the antivortex nucleates, and 
    the plus sign is for the opposite one where the antivortex ends touch the 
    surface. We consider only half of the lateral edge, otherwise the total flux would vanish.

    \begin{figure}[htpb]
        \vspace{0.5cm}
        \centering
        \includegraphics[width=0.49\textwidth]{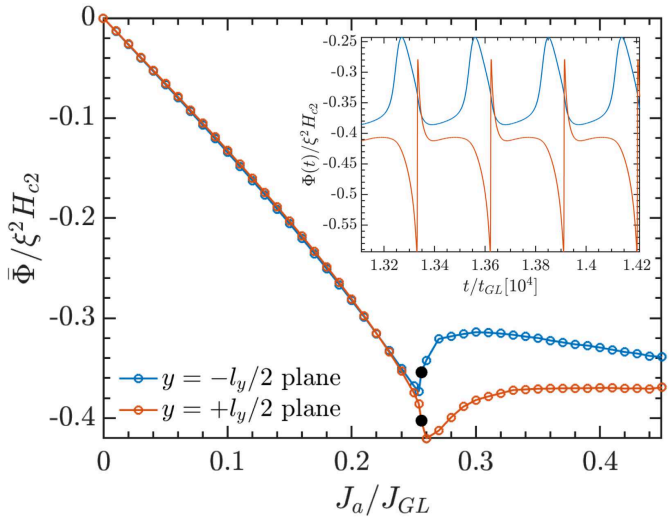}
        \caption{The main panel presents the time average of the 
        magnetic flux, $\bar{\Phi}$, across the surfaces defined 
        in Eq.~\ref{eq:eq18}; the black points 
        correspond to the beginning of the resistive state. 
        The inset shows the magnetic flux, $\Phi$, as a function 
        of time. The parameters used were $\kappa=1$, $l_z=4\xi$, 
        and $J_a=0.26J_{GL}$.
        }
        \label{fig:fig9}
    \end{figure}
    
    Fig.~\ref{fig:fig9} presents the results for the time averaged magnetic flux 
    by using the same parameters as those used in Fig.~\ref{fig:fig8}. 
    As can be clearly seen, until the transition to the resistive state, the flux 
    is approximately the same through both surfaces $y=\pm l_y/2$. 
    Nevertheless, once the resistive state sets in, they become  
    different as much as $\Delta \bar{\Phi} \approx 0.09\xi^2H_{c2} \approx 0.02\Phi_0$. 
    This is signaling that the antivortex is piercing  the $y=+l_y/2$ surface. 
    Therefore, for this to happen, the antivortex necessarily has to bend. 
    
    Since the creation and annihilation 
    of the half-closed vortex is a dynamical process, the flux evolves periodically. 
    The AC magnetic flux can be seen in the inset 
    of Fig.~\ref{fig:fig9}. The period of the AC signal depends on the applied 
    current density. For $J_a=0.26J_{GL}$ we find that the period is 
    $\tau \approx 0.03\times 10^4t_{GL}$. For low-$T_c$ materials 
    like Nb films \cite{berdiyorov2009}, 
    $t_{GL}\approx 6.72$ ps. This produces $\tau \approx 2$ ns, which 
    is in the GHz frequency range.
    
    As we can see, the measurement of the difference between the time averaged 
    magnetic flux threading at each plane, as displayed in Fig.~\ref{fig:fig9}, can 
    be an indirect method for the experimental detection of a closed vortex. Such 
    measurement is experimentally feasible by using the recently developed nanoSQUIDs 
    \cite{embon2017,anahory2020,vasyukov2013scanning}, which are 
    capable of detecting the variation of 
    the flux produced by the closed vortex in the time and length scale we used 
    in our computations.

    \begin{figure}[htpb]
        \vspace{0.5cm}
        \centering
        \includegraphics[angle=-90,width=0.49\textwidth]{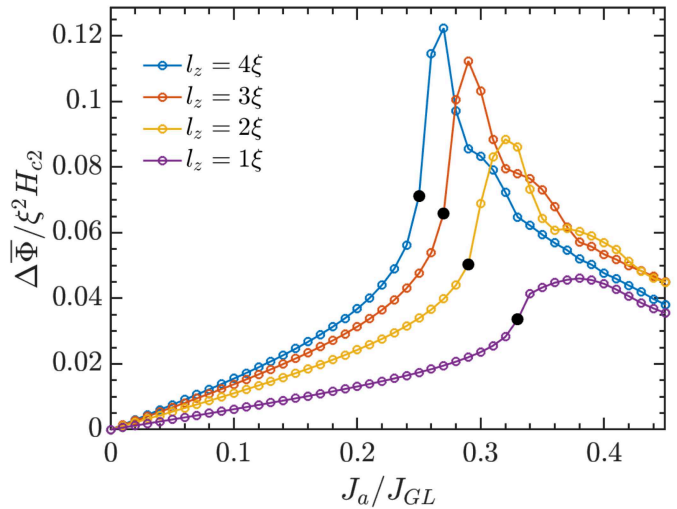}
        \caption{Difference between the average magnetic flux on both 
        sides of the sample through two vertical circuits positioned 
        on the planes $y=\pm (l_y/2+\xi)$. The domain of the circuits is given
         by $\{-\xi \le x \le \xi,\, -l_z/2 \le z \le L_z/2  \}$. 
         The value of $L_z$ was chosen such that 
         the area of the circuit above the $z=l_z/2$ surface 
         is the same for all thicknesses $l_z$. The points 
         just before the onset of the resistive state are 
         highlighted in black.
        }
        \label{fig:fig10}
    \end{figure}
    
    We must emphasize that for thin superconductors, the measurement 
    of the flux in the region prescribed in the above setup can 
    be experimentally challenging. For this reason, we also present 
    another indirect method for the detection of a closed vortex. 
    Fig~\ref{fig:fig10} shows the difference between the averaged magnetic 
    flux calculated at the lateral sides of our superconductor as 
    a function of the applied current density, for different 
    superconductor thicknesses. In contrast with the previous 
    case, here the flux is calculated from the bottom of the 
    superconductor up to a height well above the sample surface. 
    The flux is evaluated across a vertical rectangular surface, 
    located a coherence length away from the lateral surface.
    This makes the proposed experiment much more feasible.

    The black dots in Fig.~\ref{fig:fig10} represent the onset of 
    the resistive state for each thickness. We are interested in 
    current densities slightly above these values. In this region, 
    the antivortex moves through the whole sample, being expelled at the other 
    side (animations of this regime, as well as the one described below, 
    can be found in the Supplement Material \footnote{See Supplemental Material at [URL] for animations of the dynamics of the half-closed vortex.}). Due to its curvature, 
    the antivortex produces a larger magnetic 
    flux in the plane which it is moving into, increasing the flux 
    difference between each plane. Since the curvature increases 
    with the film thickness, this difference also increases with the 
    sample thickness, as shown in Fig.~\ref{fig:fig10}. Nevertheless, for large values 
    of the current density, a vortex also penetrates the superconductor 
    at the opposite side, with the pair being annihilated inside the sample, 
    reducing the impact of the curvature in the flux. The penetration of 
    this vortex becomes easier as the superconductor thickness increases, which 
    explains why the flux difference for $l_z = 4 \xi$ is smaller than for 
    $l_z = 2 \xi$ or $3 \xi$ at high current densities, for example. 
    In summary, by comparing the magnetic flux difference,  at the 
    onset of the resistive state, for films with different thicknesses, 
    we can clearly demonstrate the existence of the half-closed vortex.
    Given the inherent complexity for the direct 
    observation of a closed vortex, our indirect method brings a new 
    possibility for the first detection of such objects.
    
\section{\label{sec:sec4}Concluding remarks}

To summarize, we have shown that the combination of the 
flux lines of a vortex and an antivortex during their 
annihilation gives origin to a closed vortex loop. As we 
show here, the formation of the closed vortex depends on 
how easily the flux lines can be bent due to the action of the 
applied current, with this bending increasing with the 
superconducting film thickness and decreasing with the 
Ginzburg-Landau parameter $\kappa$. Since the motion and 
annihilation of vortices are highly dissipative processes, 
understanding their behavior is 
of fundamental importance to the design of electronic 
devices.

Our findings suggest a new 
method to experimentally observe a closed vortex. As discussed
here, closed vortices can be indirectly detected by the measuring the 
flux produced by their stray fields. We emphasize that 
the recently developed nanoSQUIDs  
are capable of performing such measurements in the time and 
length scales that our system requires.

\begin{acknowledgments}
LRC, LVT, and ES thank the Brazilian Agency FAPESP for financial 
support, grant numbers 20/03947-2, 19/24618-0, 20/10058-0, 
respectively. ES thanks Professor Alexey Samokhvalov 
for very useful discussions. ES is also grateful for the warming 
hospitality of the Department of Physics, University of Antwerp, 
where this work was finished. WAO thanks the National Council 
for Scientific and Technological Development (CNPq, Grant 309928/2018-4). 
ES thanks Professor Felipe Fernandes Fanchini for kindly 
donating two GPU cards (FAPESP, grant number 21/04655-8).
 
\end{acknowledgments}





\bibliography{main}

\end{document}